\documentclass[useAMS,usenatbib]{mnras}
\usepackage{amssymb,graphicx,natbib,longtable,hyperref}
\usepackage{breakurl}
\usepackage{tablefootnote}
\bibpunct{(}{)}{;}{a}{}{,} 
\newcommand{\oneE} {1E\,1613}
\newcommand{\oneEc} {1E\,161348--5055}
\usepackage[british]{babel}

\title[Evidence for \oneEc\ being a magnetar]{Evidence for the magnetar nature of \oneEc\ in RCW\,103}

\author[A. D'A\`i et al.]{A.~D'A\`i$^{1}$, \thanks{antonino.dai@ifc.inaf.it} 
P.\,A. Evans$^{2}$, D.\,N. Burrows$^{3}$, N.\,P.\,M. Kuin$^{4}$, 
D.\,A. Kann$^{5}$, S. Campana$^{6}$,\and A. Maselli$^{1}$, P. Romano$^{1}$,
G. Cusumano$^{1}$, V. La Parola$^{1}$,
S.\,D. Barthelmy$^{7}$,\and
A.\,P. Beardmore$^{2}$,
S.\,B. Cenko$^{7,8}$,
M. De Pasquale$^{4}$,
N. Gehrels$^{7}$,
J. Greiner$^{9}$,\and
J. A. Kennea$^{3}$,
S. Klose$^{9}$,
A. Melandri$^{6}$,
J.\,A. Nousek$^{3}$,
J.\,P. Osborne$^{2}$,
D.\,M. Palmer$^{10}$,\and
B. Sbarufatti$^{3,6}$,
P. Schady$^{9}$,
M.\,H. Siegel$^{3}$,
G. Tagliaferri$^{6}$, 
R. Yates$^{9}$,
S. Zane$^{4}$
\vspace{6pt}\\
$^{1}$ INAF/IASF Palermo, via Ugo La Malfa 153, I-90146, Palermo, Italy\\
$^{2}$ Department of Physics and Astronomy, University of Leicester, Leicester LE1 7RH, UK\\
$^{3}$ Department of Astronomy and Astrophysics, Pennsylvania State University, 525 Davey Lab, University Park, PA 16802, USA\\
$^{4}$ Mullard Space Science Laboratory, UCL, Holmbury St Mary, Dorking, Surrey, RH56NT, UK\\ 
$^{5}$ Th\"uringer Landessternwarte Tautenburg, Sternwarte 5, D-07778 Tautenburg, Germany\\
$^{6}$ INAF, Osservatorio Astronomico di Brera, via E. Bianchi 46, I-23807 Merate, Italy\\
$^{7}$ NASA Goddard Space Flight Center, Mail Code 661, Greenbelt, MD 20771, USA\\
$^{8}$ Joint Space-Science Institute, University of Maryland, College Park, MD 20742, USA\\
$^{9}$ Max-Planck-Institut f\"ur extraterrestrische Physik, Giessenbachstra{\ss}e 1, D-85748 Garching, Germany\\
$^{10}$ Los Alamos National Laboratory, B244, Los Alamos, NM 87545, USA\\
}
\date{Received 2016 July 15; accepted 2016 August 08}
\begin{document}

%\date{DRAFT VERSION}

\pagerange{\pageref{firstpage}--\pageref{lastpage}} \pubyear{0000}

\maketitle

\begin{abstract}
We report on the detection of  a bright, short, structured X-ray burst
coming from the  supernova remnant RCW\,103 on 2016 June  22 caught by
the  \emph{Swift}/Burst  Alert Telescope  (BAT)  monitor,  and on  the
follow-up   campaign    made   with    \emph{Swift}/X-Ray   Telescope,
\emph{Swift}/UV/Optical Telescope and  the optical/near infrared (NIR)
Gamma-ray   Burst    Optical   and   Near-infrared    Detector.    The
characteristics of  this flash, such  as duration and  spectral shape,
are  consistent with  typical short  bursts observed  from soft  gamma
repeaters.   The BAT  error circle  at  68 per  cent confidence  range
encloses  the point-like  X-ray source  at the  centre of  the nebula,
\oneEc. Its nature has been long  debated due to a periodicity of 6.67
h in X-rays,  which could indicate either an  extremely slow pulsating
neutron star,  or the orbital  period of  a very compact  X-ray binary
system.  We found  that 20 min before the BAT  trigger, the soft X-ray
emission of \oneEc\ was a factor of $\sim$\,100 higher than measured 2
yr  earlier, indicating  that  an outburst  had  already started.   By
comparing the spectral and timing characteristics of the source in the
2  yr before  the outburst  and  after the  BAT event,  we find  that,
besides a  change in luminosity  and spectral  shape, also the  6.67 h
pulsed profile has significantly changed with a clear phase shift with
respect to its low-flux  profile.  The UV/optical/NIR observations did
not reveal any counterpart at the  position of \oneEc.  Based on these
findings, we associate the BAT burst  with \oneEc, we classify it as a
magnetar, and  pinpoint the  6.67 h periodicity  as the  magnetar spin
period.

\end{abstract}


\section{Introduction}

RCW\,103  is  a shell  supernova  remnant  (SNR) of  $\sim$\,9  arcmin
apparent  diameter, expanding  at  around 1100  km  s$^{-1}$, with  an
estimated  age between  1350 and  3050 yr  \citep{carter97}, and  at a
distance  of  3.3  kpc   \citep{caswell75}.   \citet{frank15}  gave  a
detailed, spatially  resolved, account  of the  X-ray emission  of the
SNR, that can be modelled with an absorbed, non-equilibrium ionization
state (NEI) plane shock model, with an average temperature of 0.58 keV
and       an       absorbing        column       density       $N_{\rm
  H}$\,=\,0.95\,$\times$\,10$^{22}$ cm$^{-2}$.  The relative abundance
of the  most important  metals (Ne,  Mg, Si, S,  and Fe)  is generally
found to be  half the equivalent solar value,  reflecting the stronger
contribution  of the  metal-poor  circumstellar  medium emission  with
respect to the expected metal-rich ejecta.

The compact soft X-ray source  \oneEc\ (hereafter \oneE), which is the
neutron  star (NS)  born from  the core-collapse  supernova explosion,
lies nearly  at the centre  of the  SNR.  The association  between the
central compact object (CCO) and the  SNR is proved by a depression of
$\sim$\,1  arcmin in  the H\textsc{i}  emission of  the SNR,  which is
positionally and kinematically coincident with the location of the CCO
\citep{reynoso04}. However,  the CCO  has no confirmed  counterpart at
other wavelengths.

The X-ray  luminosity of  \oneE\ can  vary by more  than one  order of
magnitude on a  time-scale of years in  the range 10$^{33}$--10$^{35}$
erg s$^{-1}$ \citep{deluca06}.  The X-ray spectrum is rather soft, and
it can  be well described  either by the  sum of two  blackbodies with
temperatures of  0.5 keV  (and corresponding emitting  radius, $R_{\rm
  BB}$  of few  hundred  metres) and  1.0 keV  ($R_{\rm  BB}$ tens  of
metres), respectively, or  by the sum of a soft  black-body of 0.5 keV
and  a  steep power-law  of  photon  index $\sim$\,3  \citep{deluca06,
  esposito11}.  One of its most enigmatic features is a periodicity of
6.67 h  found in  a long \emph{XMM-Newton}  observation of  the source
\citep{deluca06}.   It is  debated if  the periodicity  refers to  the
rotational period of an extremely slow,  and peculiar, NS, or it is an
orbital modulation  of an accreting  compact X-ray binary  system.  In
the first hypothesis the NS should  have an extreme magnetic field ($B
\sim$ 10$^{13}$--10$^{15}$ G) as is  typical of the so-called magnetar
systems. This could possibly explain the very long spin period because
of a large spin-down due to  the interaction with a fossil disc formed
from the  debris of the supernova  (SN) explosion. In the  latter case
the system would be a quite odd example of a very young low-mass X-ray
binary system, even if the requirement of an extreme magnetic field is
probably    still   needed    \citep{deluca06,   li07,    pizzolato08,
  bhadkamkar09, ikhsanov13}.

In this  paper, we report  on the recent  discovery of a  bright X-ray
flash  observed on  2016  June 22  with  the \emph{Swift}/Burst  Alert
Telescope (BAT) instrument \citep{gcn19547}  from the RCW\,103 region.
The position of the hard X-ray source responsible for the X-ray flash,
labelled   SGR  1617-5103,   is  compatible   with  the   position  of
\oneE\ \citep{atel9180, atel9183}. We  present a detailed spectral and
timing study of the X-ray emission  of \oneE\ before and after the BAT
trigger, finding the CCO in an  outburst state.  We also report on the
search for  a transient UV/optical/NIR counterpart  with the Gamma-Ray
Optical       and      Near-infrared       Detector      \citep[GROND;
][]{Schady2016ATel9184}  and \emph{Swift}/UV/Optical  Telescope (UVOT)
to the outbursting  source, finding only upper limits  at the position
of  the CCO.   We propose  the identification  of SGR  1617--5103 with
\oneE\  and  we  discuss  the   implications  of  this  discovery  for
constraining the nature of \oneE.

\section{OBSERVATIONS, DATA REDUCTION, AND ANALYSIS}

The \emph{Swift} \citep{gehrels04} satellite was launched in 2004 with
the  primary  goal  of  detecting gamma-ray  bursts.   The  scientific
payload  of \emph{Swift}  comprises three  different instruments:  BAT
\citep{barthelmy05}, a  coded mask  telescope sensitive to  photons in
the  15--350 keV  range and  with  a 2  sr  field of  view, the  X-ray
Telescope \citep[XRT;][]{burrows05}, that covers  the soft X-ray range
(0.3--10  keV)  with a  field  of  view  $\sim$\,23.6 arcmin  and  the
UV/Optical  telescope \citep[UVOT;][]{roming05},  that with  different
filters can  cover the 1270--6240  \textrm{\AA} wavelength range  on a
17\,$\times$\,17  arcmin$^{2}$  field   of  view.   \emph{Swift}/X-Ray
Telescope (XRT) can perform X-ray  observations in imaging mode with a
2.5 s frame-time (Photon Counting mode, PC) or in Windowed Timing (WT)
mode at a higher timing resolution of 1.8 ms.

We performed data extraction  and reduction using the \textsc{heasoft}
software (v.  6.19)  developed and maintained by the  NASA High Energy
Astrophysics  Science Archive  Research Center  (\textsc{heasarc}). We
used  \textsc{xspec} v.\,12.9.0  for spectral  analysis.  Spectra  are
re-binned to a minimum of 20 counts per energy channel to allow use of
$\chi^2$  statistics.   Errors on  spectral  parameters  are given  at
90\,\% confidence level ($\Delta \chi^2$\,=\,2.706).  Luminosities are
given assuming isotropic emission and a distance of 3.3 kpc.

\subsection{BAT analysis and results} \label{sect:bat}
\begin{figure}
\begin{center}
\includegraphics[width=\columnwidth]{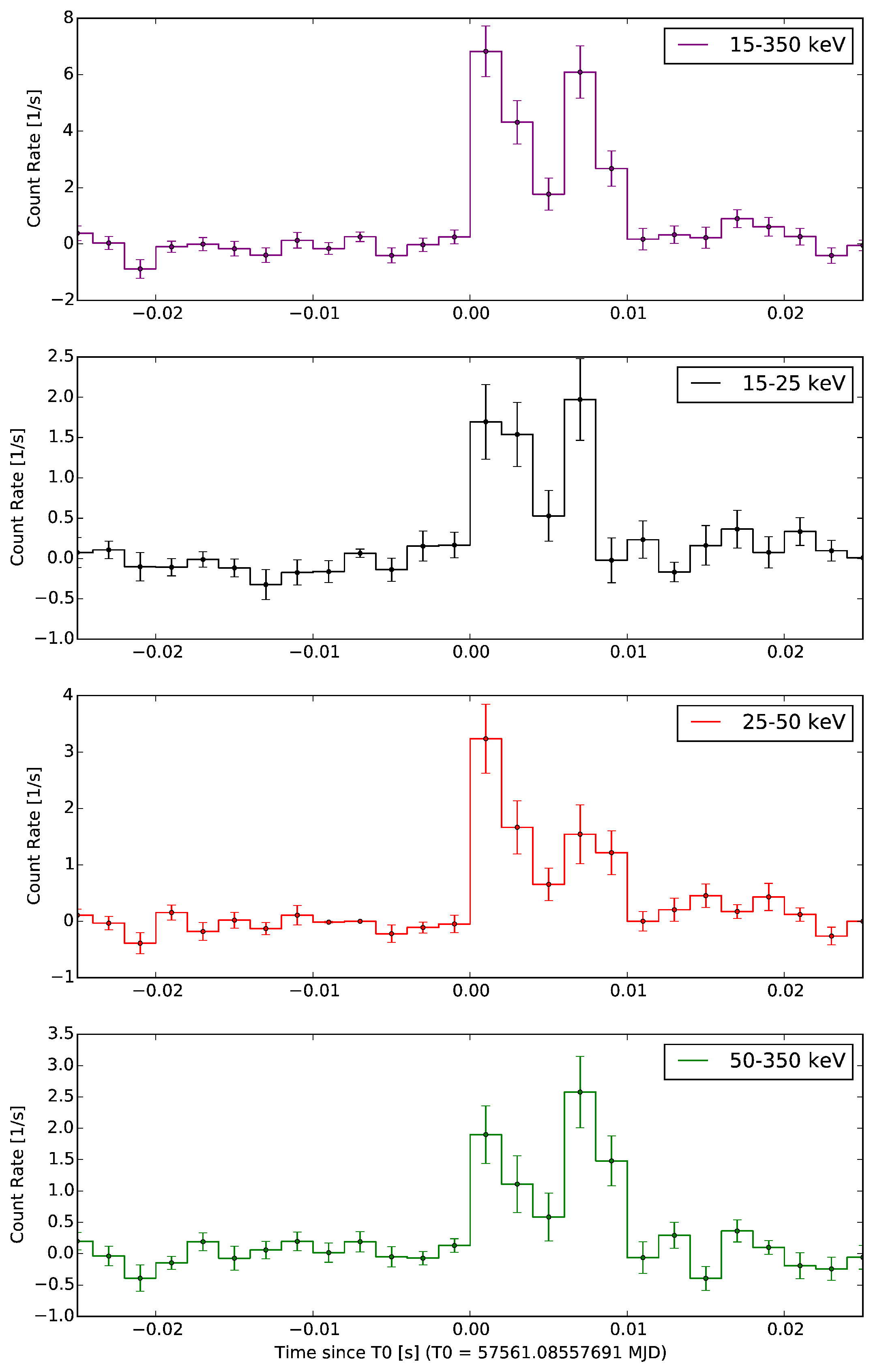}
\end{center}
\caption{\emph{Swift}/BAT  mask-weighted  light  curves in different energy bands.}
\label{fig:BAT_lc}
\end{figure}

BAT triggered  on SGR 1617--5103  at 02:03:13.845 \textsc{ut}  on 2016
June 22  ($T_0$\,=\,57561.08557691 MJD).  The BAT  mask-weighted light
curve  shows a  structured short  pulse with  $T_{90}$ (the  time over
which a burst emits from 5\% of  its total measured counts to 95\%) of
8.0\,$\pm$\,4.5 ms. In Fig.~\ref{fig:BAT_lc} we show the burst profile
in some  energy-selected bands  with a  time resolution  of 2  ms. The
profile is consistent with a  single double-peaked burst or, possibly,
with two consecutive, very close, shorter bursts.

\begin{figure}
\begin{center}
\includegraphics[width=\columnwidth]{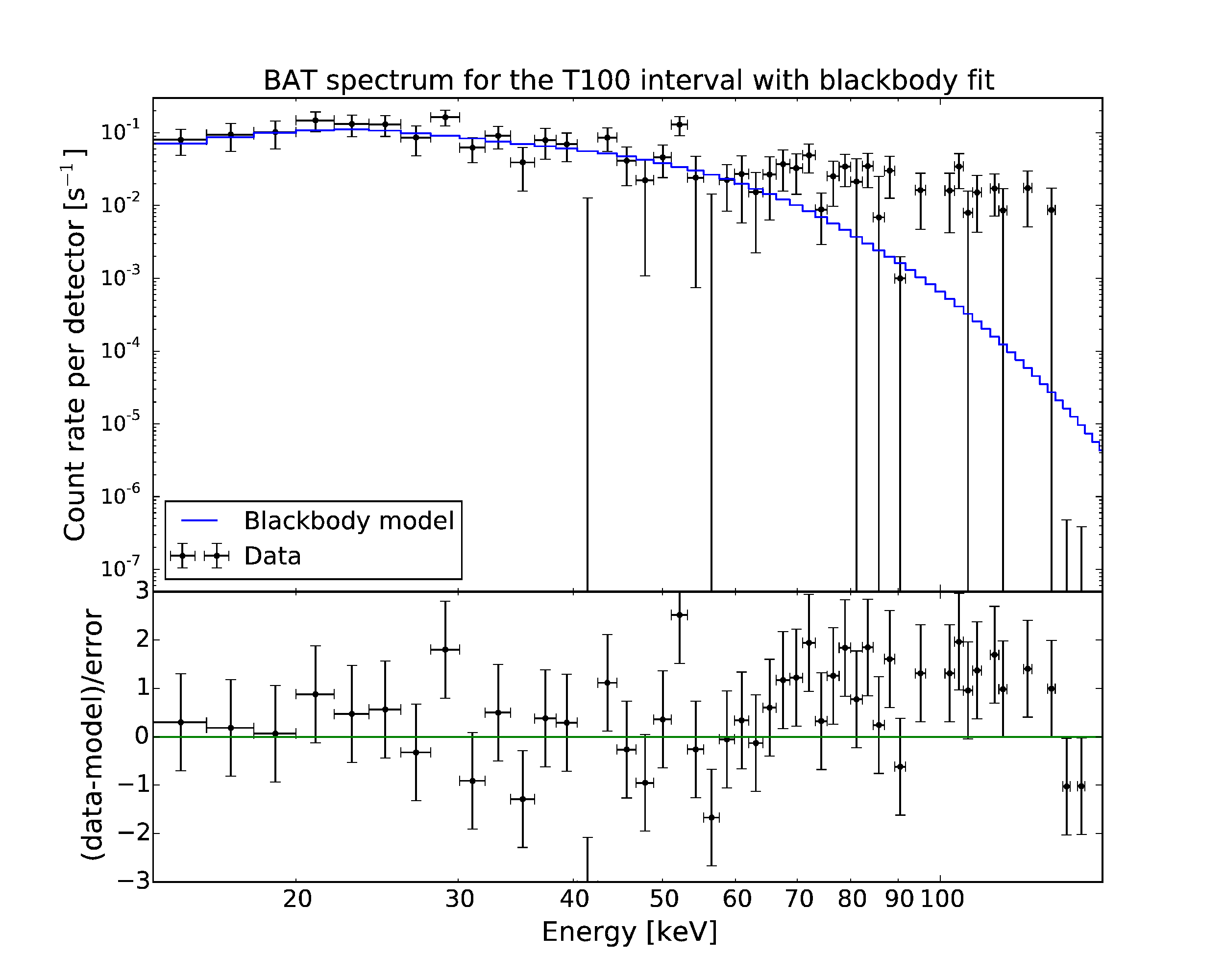}
\end{center}
\caption{15--150  keV  BAT  data, blackbody  best-fitting  model  and
  residuals  in   units  of  $\sigma$  for   the  time-averaged  burst
  emission.}
\label{fig:BAT_spec}
\end{figure}

We fitted  the time-averaged  BAT spectrum from  $T_0$ to  $T_0$+10 ms
using a  power-law model.  The  spectral fit shows a  power-law photon
index   of  2.38\,$\pm$\,0.21   and   a  fluence   (15--150  keV)   of
$(1.25\,\pm\,0.22) \times 10^{-8} \ \rm erg \ cm^{-2}$, with a reduced
chi-squared ($\chi_{\rm  red}^2$) of  1.55 for  57 degrees  of freedom
(dof).  A blackbody model  provides a significantly better description
of  the  data,   with  a  $\chi_{\rm{red}}^2$  of  1.18   for  57  dof
(Fig.~\ref{fig:BAT_spec}).      The    blackbody     temperature    is
$kT$\,=\,10.3\,$\pm$\,1.3 keV, the blackbody radius is 1.2\,$\pm$\,0.2
km, and  the corresponding fluence (15--150  keV) is 1.51\,$\pm$\,0.24
$\times$ 10$^{-8}$ erg  cm$^{-2}$, which translates for  a distance of
3.3 kpc  into an isotropic  total energy release  of (2.0\,$\pm$\,0.3)
$\times$\,10$^{37}$ erg.  A time-resolved fit from $T_0$ to $T_0$+5 ms
and from $T_0$+5 to $T_0$+10 ms, is consistent for both intervals with
a thermal blackbody emission of temperature $kT$\,=\,8.7$\pm$\,1.7 and
$kT$\,=\,10$_{-3}^{+5}$     keV,     and    a     blackbody     radius
1.7$^{+0.9}_{-0.5}$  and  0.9$^{+0.9}_{-0.5}$  km for  the  first  and
second peak, respectively.

The BAT  ground-calculated position is RA,  Dec.  = 244.$^{\circ}$385,
--51.$^{\circ}$047, with  an uncertainty of  2.0, 1.7, and  1.2 arcmin
for 95,  90 and  68 per cent  containment, respectively.   The partial
coding was 71 per cent.

\subsection{XRT analysis and results} \label{sect:xrt}

\emph{Swift}/XRT observed  RCW 103 both in  PC and WT modes.   In this
work we will focus only on  the X-ray emission coming from the central
object  inside the  RCW 103  nebula,  so we  extracted the  high-level
scientific  products using  a circular  region of  20-pixel radius  (1
pixel\,=\,2.36  arcsec) at  the  most accurate  coordinates of  \oneE~
\citep[RA\,=\,244.400958,  Dec.\,=\,--51.040167;][]{deluca08}.  Photon
arrival  times were  corrected to  the Solar  system barycentre  using
these coordinates  with the \textsc{barycorr} tool.   For observations
in PC  mode, when the  count rate  inside this region  was $\geq$\,0.5
counts s$^{-1}$ (cps), we adopted an annular extraction region of 3--5
pixels inner radius, depending on  the source brightness, and 20 pixel
outer radius to  take into account pile-up effects.   Light curves are
corrected for vignetting, bad columns, and point spread function (PSF)
extraction regions using the \textsc{xrtlccorr} tool.

\emph{Swift}/XRT has regularly observed the field of RCW 103 since the
start of  the mission.   \citet{esposito11} examined  the observations
spanning the  period 2006--2011,  finding that  \oneE~ had  an average
1--10  keV  flux  of  $\sim$\,1.7 $\times$  10$^{-12}$  erg  cm$^{-2}$
s$^{-1}$, modulated at the 6.675 h  (24030 s) period.  For the aims of
the present  work, we examined  all observations performed  after 2014
January,  in order  to have  a statistically  robust benchmark  of the
\textit{quiescent}   state  of   \oneE\   before   the  BAT   trigger.
\emph{Swift} monitored the  RCW 103 nebula with a visit  of few ks per
month till  2016 June,  while starting  from 2016  June 22,  after the
detection  of  the BAT  X-ray  burst,  \emph{Swift} began  an  intense
observing  campaign,  collecting a  total  of  63.4 ks  exposure  time
between   2016  June   22   and   2016  July   12.    We  present   in
Table~\ref{obs-log}  a summary  of all  the observations  analysed for
this work.

\begin{figure*}
\centering
\includegraphics[width=17cm]{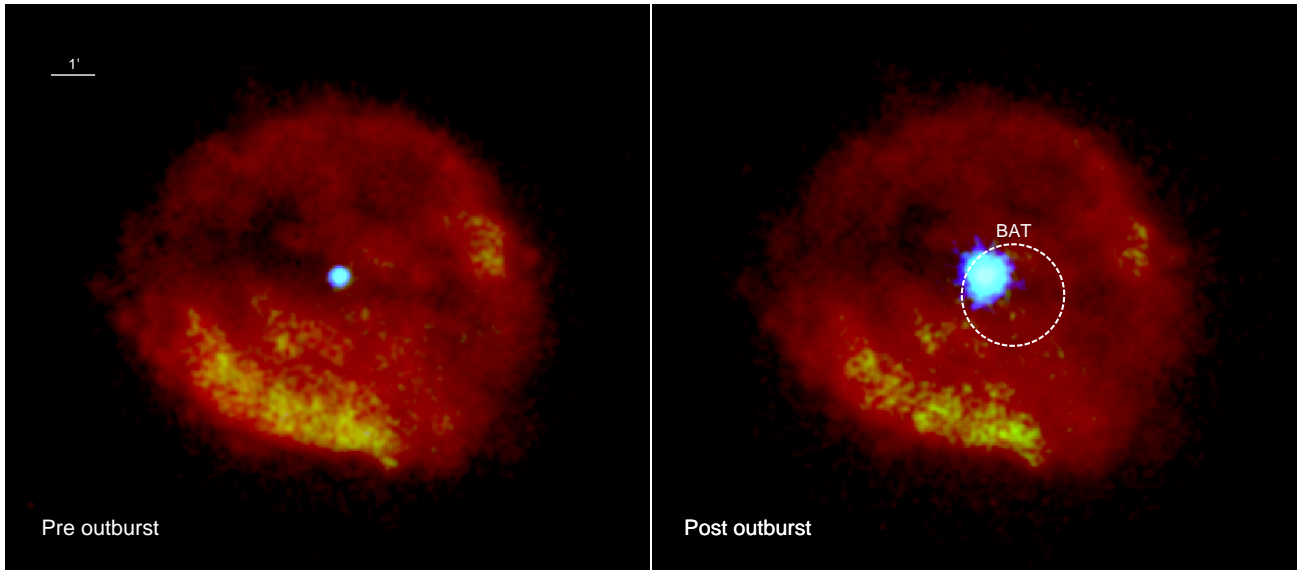}
\caption{\emph{Swift}/XRT  stacked image  of all  observations of  the
  SNR, RCW\,103, in the 2 yr before the outburst (left-hand panel) and
  after 2016 June  22 (right-hand panel).  The total  exposure time is
  45 ks in the left-hand image and 57 ks in the right-hand one. Photon
  energy is colour  coded for different energy bands:  red, green, and
  blue correspond  to the  0.5--1.5, 1.5--2.5,  and the  2.5--10.0 keV
  bands, respectively. The white dashed  line in the right panel marks
  the  68 per  cent  confidence level  of  the \emph{Swift}/BAT  error
  circle (radius 1.2 arcmin; see the  online version of this paper for
  the colour figure).}
\label{fig:image}
\end{figure*}

In Fig.~\ref{fig:image} we  show a false-colour image of  the field of
RCW 103 in three energy bands (0.5--1.5, 1.5--2.5, and 2.5--10.0 keV),
obtained   by    stacking   all   the   PC    mode   observations   of
Table~\ref{obs-log}  before (left-hand  panel)  and after  (right-hand
panel)  the BAT  trigger.  The  right-hand image  shows the  BAT error
circle, at the 68\% confidence level, of the short X-ray burst of June
22, that is  consistent with the position of \oneE.   It is clear that
this source  brightened after June  22, whereas no  other \textit{new}
point-like  source, and  no extended  brightening of  any part  of the
nebula  appears as  a  possible, alternative,  soft X-ray  counterpart
candidate of the BAT burst.

\begin{figure}
\centering
\includegraphics[width=\columnwidth]{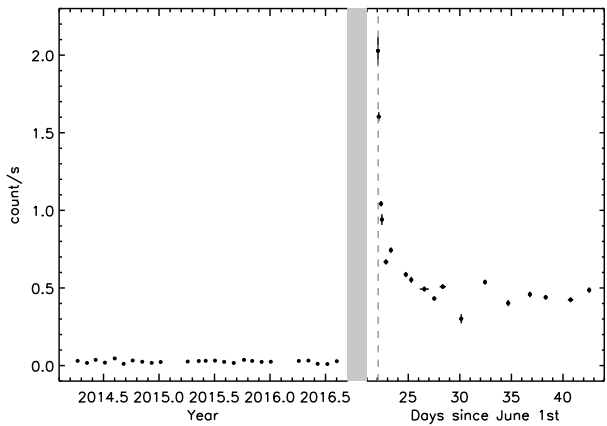}
\caption{Light  curve  (2--10  keV   range)  of  the  \emph{Swift}/XRT
  observations performed between 2014 January 15 and 2016 July 12. The
  grey-shaded line indicates a change  in the $X$-axis scale, the dotted
  line indicates the time of the BAT burst.}
\label{fig1lc}
\end{figure}

We show the  2--10 keV \oneE\ light curve  in Fig.~\ref{fig1lc}, where
each  point  is  the  time-averaged  count  rate  in  one  observation
identification number (ObsID, see Table~\ref{obs-log}).  In the period
2014 January  -- 2016 May the  source flux remained at  an approximate
constant   low  flux   level  with   an  observed   average  flux   of
2.7\,$\times$\,10$^{-12}$ erg cm$^{2}$ s$^{-1}$  for the entire period
(the  unabsorbed   flux  is  8.2\,$\times$\,10$^{-12}$   erg  cm$^{2}$
s$^{-1}$).   It  is important  to  note  that the  observation,  ObsID
00030389032, was performed about 20  min before BAT detected the short
burst, and in this observation the source was already about two orders
of magnitude brighter  than in all previous  observations (the closest
being  the   ObsID  00030389030  performed   on  May  16).    All  the
observations performed after June 22 clearly show \oneE\ in a brighter
state, confirming an  ongoing outburst state.  Both  the quiescent and
the active  state emission  are modulated at  the 6.67  hr periodicity
(see  Section~\ref{sec:timing}),  which  accounts   for  most  of  the
variance   observed  in   all   the  ObsIDs   (see   last  column   in
Table~\ref{obs-log}). However,  we note that the  average rates during
June 22 are  significantly higher with respect to  the following days,
and they are not due to periodic modulation but imply an intrinsically
brighter state. The  light curve suggests an  almost exponential decay
on the day BAT  detected the burst, and a fit of  the light curve with
observations performed between June 22 and  23 results in a decay time
of 1.3\,$\pm$\,0.3 d.

\begin{figure}
\centering
\includegraphics[height=\columnwidth, angle=-90]{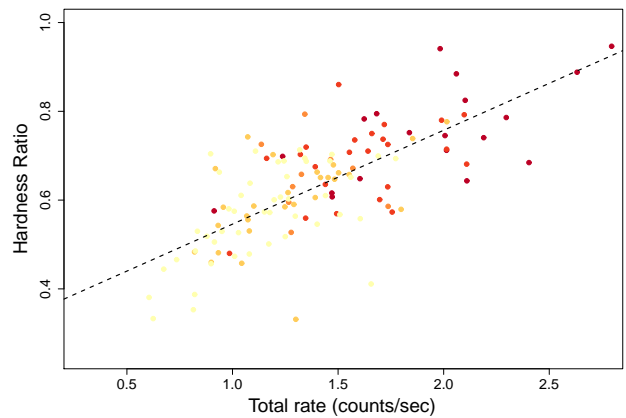}
\caption{Hardness ratio  (3--10 keV/1--3 keV  rates) as a  function of
  the total rate for PC mode observations of the outburst.  Each point
  is  the  averaged rate  in  1  ks.  Error  bars  are  not shown  for
  clarity's sake,  a point  with typical  error bars  is shown  in the
  left-hand corner of the figure. Time  is colour coded going from red
  (first  observations) to  yellow  (last  observations).  The  dashed
  black line shows the best-fitting linear regression line. }
\label{fig:hardness}
\end{figure}

We then extracted two energy-filtered light curves in the 1.0--3.0 and
3.0--10.0 keV bands using all PC  mode observations with a bin size of
1 ks after the  burst event to study the spectral  hardness ratio as a
function of the total  rate and time. We found that  the total rate is
significantly   correlated  with   the   spectral  hardness   (Pearson
correlation  coefficient 0.7\,$\pm$\,0.1  at  95  per cent  confidence
level), while there is insignificant  evolution as a function of time,
besides  the  relative  higher  brightness  and  corresponding  higher
hardness ratio  values in the  first day of  the outburst as  shown in
Fig.~\ref{fig:hardness}.
  
%}

\begin{table*} \centering
\begin{tabular}{l crrr}
\hline
ObsID (mode)             &  $T_{\rm start}$--$T_{\rm stop}$ & Exposure & Mean rate &  $\sigma$\phantom{aa} \\
                         &     (\textsc{tt}) & (s)        &  \phantom{aa}(10$^{-2}$ cps) & (cps)   \\
                         \hline\noalign{\smallskip}
                      &   \multicolumn{4}{c}{Pre-outburst observations} \\
00030389004--00030389030  & 2014-01-15 07:17:00 2016-05-16 15:37:54 &  44480      & 2.9 & 3.2 \\
\noalign{\smallskip}
                      &   \multicolumn{4}{c}{June 22 observations} \\
00030389032 (PC)          & 2016-06-22 01:31:47--2016-06-22 01:42:54 &  644        & 204 & 34        \\
00700791000 (WT)          & 2016-06-22 02:05:00--2016-06-22 03:40:49 & 2222        & 161 & 27 \\
00700791001 (WT)          & 2016-06-22 04:37:34--2016-06-22 12:46:52 & 4278        & 101 & 32  \\
00700791001 (PC)          & 2016-06-22 08:14:10--2016-06-22 13:15:26 & 2163        & 95 &  21  \\
00700791002 (PC)          & 2016-06-22 14:31:46--2016-06-23 01:49:31 & 6398        & 68 &  28  \\
\noalign{\smallskip}
                      &   \multicolumn{4}{c}{Late-time outburst observations} \\

00700791003 (PC)          & 2016-06-23 03:34:40--2016-06-23 11:35:35 & 4932        & 73 & 26   \\
00700791004 (PC)          & 2016-06-24 15:55:27--2016-06-24 20:42:54 & 4260        & 60 & 29  \\
00700791005 (PC)          & 2016-06-25 04:29:35--2016-06-25 09:27:27 & 3551        & 56 & 18 \\
00700791006 (PC)          & 2016-06-26 02:48:57--2016-06-27 00:01:53 & 3529        & 50 & 15   \\
00700791007 (PC)          & 2016-06-27 10:47:59--2016-06-27 14:26:54 & 4924        & 43 & 18  \\
00700791008 (PC)          & 2016-06-28 01:03:52--2016-06-28 15:32:13 & 4652        & 51 & 24   \\
00700791009 (PC)          & 2016-06-30 02:29:07--2016-06-30 04:04:34 &  737        & 29 & 12  \\
00700791010 (PC)          & 2016-07-02 07:23:01--2016-07-02 13:51:53 & 4472        & 55 & 19  \\
00700791011 (PC)          & 2016-07-04 13:24:56--2016-07-04 19:40:44 & 2090        & 41 & 21  \\
00700791012 (PC)          & 2016-07-06 14:42:44--2016-07-06 23:15:54 & 3079        & 46 & 16  \\
00700791013 (PC)          & 2016-07-08 05:12:33--2016-07-08 10:13:53 & 4170        & 44 & 17  \\
00700791014 (PC)          & 2016-07-10 11:13:18--2016-07-11 00:01:53 & 3901        & 45 & 20  \\
00700791015 (PC)          & 2016-07-12 09:33:02--2016-07-12 06:27:54 & 3421        & 50 & 23   \\
\hline 
\end{tabular}
\caption{\label{obs-log}  Log  of the  \emph{Swift}/XRT  observations.
  The  columns  show  the  identification  code  of  each  observation
  (ObsID), the $T_{\rm start}$ and $T_{\rm stop}$ times in Terrestrial
  Time  (\textsc{tt},   \textsc{ut}\,=\,\textsc{tt}+16.54836  s),  the
  ObsID exposure, the 2--10 keV mean count rate in the 20-pixel region
  centred on the source, and  the standard deviation ($\sigma$) with a
  time  bin of  50  s.  For WT  observations,  because  of the  higher
  background from  the nebular emission  in the collapsed  rows, rates
  are over-estimated by $\sim$\,10 per  cent with respect to the rates
  observed in PC mode.}
\end{table*}

We  extracted a  time-averaged spectrum  for all  observations between
2014 January  15 and 2016  May 16 (\textit{quiescent  spectrum}, PRE),
and a  time-averaged spectrum for  all the PC mode  observations after
June  22 (\textit{late-time  outburst spectrum},  POST).  We  verified
that no  statistically significant spectral change  occurred along the
3-week  long  \emph{Swift}/XRT  follow-up   on  a  week-averaged  time
interval. We  studied observations  taken on  June 22  individually to
determine   short-term  intrinsic   rapid   spectral  and   luminosity
variations.

Because \oneE\ is embedded in the  SNR nebula, we modelled the nebular
emission as an additive component  to the total spectrum, adopting the
\textsc{pshock}    component     following    \citet{frank15}.     The
\textsc{pshock} model is a  constant temperature, plane-parallel shock
plasma model,  characterized by  the following parameters:  the plasma
temperature  ($kT_{\rm plasma}$),  the abundance  of the  most diffuse
elements  (Abund),  the  lower  and  upper  limit  on  the  ionization
time-scale    ($\tau_{\rm    l}$    and    $\tau_{\rm    u}$,    where
$\tau$\,=\,$n_{\rm e}  t$ and $n_{\rm  e}$ is the electron  density in
the post-shock  region) and a normalization  value, $N_{\rm p}$=EM/$(4
\pi D_{\rm  A}^2)\,\times\,10^{-14}$, related to the  angular diameter
distance to the source expressed in  cm ($D_{\rm A}$) and the emission
measure (EM=$\int  n_{\rm e}  n_{\rm H}  {\rm d}V$).   In our  fits we
froze the $\tau_{\rm l}$ value  to its default ($\tau_{\rm l}$\,=\,0).
Finally, we choose a background  region outside the RCW\,103 nebula in
a zone free of other contaminating sources.

\begin{figure*}
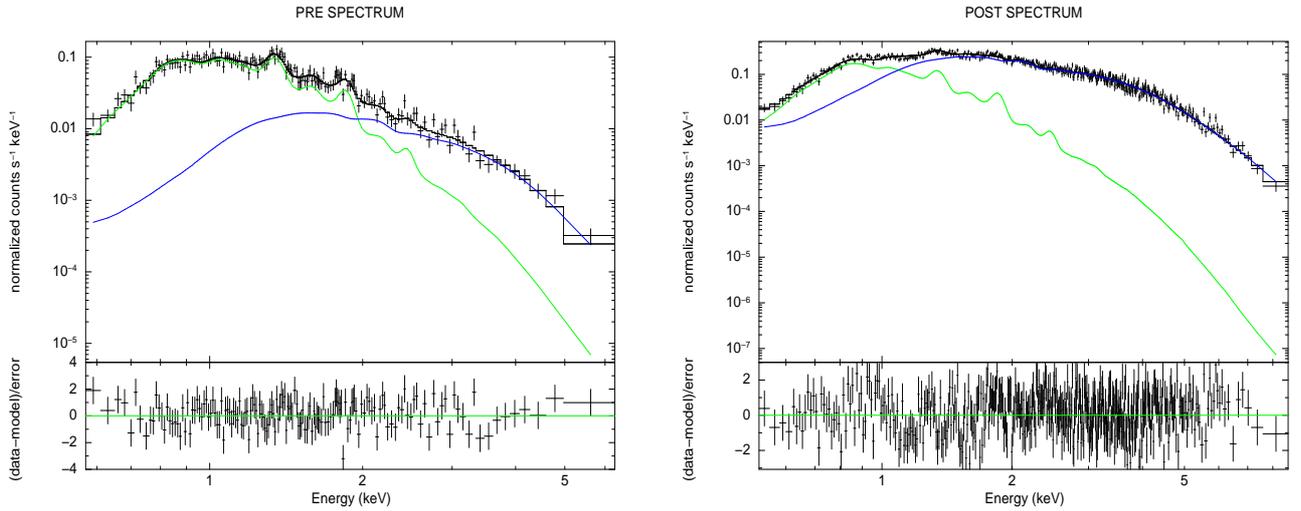

\centering
\begin{tabular}{cc}
\includegraphics[height=\columnwidth, width=0.8\columnwidth,  angle=-90]{fig6.ps}&
\includegraphics[height=\columnwidth, width=0.8\columnwidth,  angle=-90]{fig7.ps}\\
\end{tabular}
\caption{Comparison  of  the  time-averaged data,  spectral  shape  of
  \oneE, best-fitting  model and residuals  in units of  $\sigma$ from
  the  best-fits   shown  in  Table~\ref{tab:specfits}.   Left  panel:
  pre-outburst  spectrum.  Right  panel: late-time  outburst spectrum.
  We  show in  green the  \textsc{pshock} component  arising from  the
  nebular emission, and in blue the harder, thermal component from the
  compact source.}
\label{fig:xrtspec}
\end{figure*}

For the quiescent spectrum, the  emission from \oneE\ is modelled with
a blackbody.  Both the nebular  and compact components are absorbed by
interstellar  matter [we  used \textsc{phabs},  with \citet{asplund09}
  abundances and \citet{verner96} cross-sections].  The values for the
nebular  emission   are  consistent   with  the  values   reported  in
\citet{frank15}, while  the compact  emission is  well described  by a
soft  thermal component  of 0.58\,$\pm$\,0.05  keV temperature,  whose
corresponding emitting radius is 0.44  km.  The $\chi_{\rm red}^2$ for
this  model  is  0.98  (175 dof).   We  report  spectral  best-fitting
parameters    and    errors    in     the    left-hand    column    of
Table~\ref{tab:specfits} and data, best-fitting model and residuals of
this  best  fit  in  the left  panel  of  Fig.~\ref{fig:xrtspec}.   We
verified that  this spectral decomposition is  solid against different
extraction  regions. By  taking different  extraction radii,  the only
spectral parameter that must be allowed to vary to obtain satisfactory
fits  ($\chi^2_{\rm   red}$  $\approx$   1)  is   the  \textsc{pshock}
normalization value.

With respect to the quiescent spectrum, the late-time (from June 23 to
July  12)  time-averaged outburst  spectrum  shows  a change  both  in
luminosity and  in spectral shape,  the most evident aspect  being the
presence of  an excess of  flux at  harder X-ray energies.   To better
constrain these changes, we performed first a fit of the post-outburst
spectrum fixing  the values  of the  \textsc{pshock} component  to the
best-fitting values of the pre-outburst spectrum; in a second step, we
made a  combined fit of  the pre-  and post-outburst spectra  with the
values  of the  \textsc{pshock} component  tied together  for the  two
spectra.  In both cases,  we allowed the \textsc{pshock} normalization
free to  vary to take  into account the different  extraction regions.
We assumed that  the nebular emission is not able  to promptly respond
and be modified by the CCO X-ray outburst, also because the extraction
radius  used to  build the  source spectrum  is 2.5  light years  at a
distance of 3.3  kpc.  We then modelled the  post-outburst hard excess
component,  applying to  the  compact thermal  emission a  convolution
Comptonized  kernel  \citep[\textsc{simpl}  model  in  \textsc{xspec},
][]{steiner09}, that  upscatters a  fraction of the  blackbody photons
(\textsc{simpl} $f_{\rm  sc}$ parameter)  into a  power law  of photon
index   $\Gamma$.    The   luminosity    of   the   compact   emission
(\textsc{simpl*bbody} component)  increased by a factor  of $\sim$\,20
(0.5--10  keV range),  while the  blackbody temperature  is consistent
with its  pre-outburst value.  About  10 per  cent of the  photons are
up-scattered, giving  rise to  the hard X-ray  excess.  We  report the
spectral best-fitting parameters and errors  for the different fits in
Table~\ref{tab:specfits} and data, best-fitting model and residuals of
the pre- and post-outburst spectra in Fig.~\ref{fig:xrtspec}.  We note
that  both   methods,  with  frozen,  or   with  tied  \textsc{pshock}
parameters, lead to very similar results.
 
Alternatively, a statistically similar fit can be obtained by removing
the \textsc{simpl} component and adding a  power law. In this case, we
obtained  a  best-fitting  photon   index  of  0.9$_{-0.9}^{+0.6}$,  a
power-law  luminosity  of  0.77\,$\pm$\,0.13\,$\times$\,10$^{34}$  erg
s$^{-1}$,  while  the  \textsc{bbody}  $kT_{\rm  BB}$  temperature  is
0.64\,$\pm$\,0.10 keV and its luminosity is (5.7\,$\pm$\,0.3) $\times$
10$^{34}$ erg s$^{-1}$.

\begin{table*}
\begin{tabular}{llll|cc}
\hline
      Parameter     & Units                      & PRE                       & POST                    & COMBINED FIT (PRE)  & COMBINED FIT (POST) \\ \hline

      $N_{\rm H}$ & 10$^{22}$ cm$^{-2}$           & 1.23$_{-0.10}^{+0.15}$    & 1.03\,$\pm$\,0.06    & \multicolumn{2}{c}{1.19$_{-0.08}^{+0.28}$} \\

\textsc{pshock} $kT_{\rm plasma}$  & keV          & 0.58\,$\pm$\,0.12         & 0.58                   & \multicolumn{2}{c}{0.58$_{-0.23}^{+0.07}$}\\       
\textsc{pshock} $Abund$ &                         & 0.47$_{-0.11}^{+0.18}$    & 0.47                   & \multicolumn{2}{c}{0.7$_{-0.2}^{+0.4}$}\\
\textsc{pshock} $\tau_u$ &10$^{11}$  s cm$^{-3}$  & 2.4$_{-1.1}^{+1.9}$       & 2.4                    & \multicolumn{2}{c}{2.2$_{-0.7}^{+1.1}$}\\
\textsc{pshock} $N_p$    &10$^{-2}$               & 2.428$_{-0.007}^{+0.020}$ & 2.439$\pm$0.004        & 1.6$_{-0.6}^{+5}$ & 2.6\,$\pm$\,0.1\\
\textsc{pshock} Flux & 10$^{-11}$ erg s$^{-1}$    & 4.5\,$\pm$\,1.2           & 4.7\,$\pm$\,1.0        & 4.4\,$\pm$\,0.2 &  4.9\,$\pm$\,0.9\\

\textsc{simpl} $\Gamma$ &                         &                          & 2.5\,$\pm$\,1.5         &  &2.6\,$\pm$\,1.5\\
\textsc{simpl} $f_{\rm scat}$ &                   &                          & 0.11$_{-0.04}^{+0.16}$  &  &0.12$_{-0.04}^{+0.16}$\\

\textsc{bbody} $kT_{\rm BB}$ & keV            & 0.58\,$\pm$\,0.05             & 0.62$_{-0.03}^{+0.02}$       & 0.52\,$\pm$\,0.05 & 0.62$_{-0.05}^{+0.02}$ \\
\textsc{bbody} $R_{\rm BB}$ & km              & 0.44\,$\pm$\,0.15             & 1.84\,$\pm$\,0.15          & 0.6\,$\pm$\,0.2   & 1.84\,$\pm$\,0.20 \\
\textsc{bbody} $L$ & 10$^{34}$ erg s$^{-1}$   & 0.27\,$\pm$\,0.09             & 6.5\,$\pm$\,0.2            & 0.35\,$\pm$\,0.1  & 6.3\,$\pm$\,0.1 \\

\textsc{simpl*bbody} Flux & 10$^{-11}$ erg s$^{-1}$    &                   &   5.0\,$\pm$\,0.1             &                   &  5.1\,$\pm$\,0.1\\

$\chi^2$/dof         &                            & 171/175               & 492/410                    & \multicolumn{2}{c}{680 / 586} \\
\hline
\end{tabular}
\caption{\label{tab:specfits} Spectral best-fitting  parameters of the
  time-averaged  emission of  \oneE\  and of  its surrounding  nebula,
  before  (PRE) and  after  (POST) 2016  June 22.  The  fluxes of  the
  different components are  calculated in the 0.5--10  keV range using
  the \textsc{cflux} component. In the POST column the \textsc{pshock}
  parameter values  are frozen  to the corresponding  best-fitting PRE
  values.  In  the COMBINED FIT, the  PRE and POST spectra  are fitted
  together, but the \textsc{pshock} parameters are tied.}
\end{table*}

On June  22, \emph{Swift} had  probably observed \oneE\ very  close to
the start, or  the peak, of its outburst and  observations on this day
also  registered  the  highest  luminosity  values  during  the  whole
outburst.    In    Table~\ref{tab:specfits-22june},   we    show   the
best-fitting spectral results for each  ObsID, adopting the same model
of  the  late-time  outburst  spectrum with  the  \textsc{pshock}  and
$N_{\rm  H}$  values  frozen  to   the  POST  best-fitting  values  of
Table~\ref{tab:specfits}.

Because of  the low statistics, the  fit could not well  constrain all
the parameters, where  the main uncertainties affect the  shape of the
hard X-ray tail.  However, there is  a clear indication that the X-ray
emission softened passing from  ObsID 00030389032 to ObsID 00700791000
and     the    source     luminosity     rapidly    decreased     from
$\sim$\,3\,$\times$\,10$^{35}$  to  $\sim$\,1 $\times$  10$^{35}$  erg
s$^{-1}$ in the following observations.
 
We note  that, although  \emph{Swift}/XRT pointed towards  \oneE\ just
$\sim$\,100 s after the burst, it  collected data for only few seconds
and, because of an Earth limb constraint, almost all the data for this
ObsID were taken at a later time starting at 03:04:46 \textsc{ut} June
22, about 1  h after the burst, so that  this luminosity drop happened
in less than 80 min, in a temporal window where BAT detected the X-ray
burst.

\begin{table*}
\begin{tabular}{ll llll}
\hline
      Parameter     & Units                     & 00030389032                       & 00700791000 & 00700791001 & 00700791002 \\ \hline

\textsc{simpl} $\Gamma$ &                        & 1.4$_{-0.2}^{+1.3}$    & 3.1$_{-0.3}^{+0.7}$ & 3.9$_{-1.7}^{+0.9}$ & 1.15$_{-0.11}^{+1.7}$   \\
\textsc{simpl} $f_{\rm sc}$ &                        & $>$0.55                & $>$0.53             &   $>$0.18           & 0.41$_{-0.18}^{+0.12}$ \\

\textsc{bbody} $kT_{\rm{BB}}$ & keV              &  0.69$_{-0.09}^{+0.20}$  & 0.58$_{-0.08}^{+0.16}$ & 0.58$_{-0.03}^{+0.09}$ & 0.60$_{-0.07}^{+0.02}$ \\

\textsc{bbody} $R_{\rm{BB}}$ & km                &  3.1\,$\pm$\,2.5          & 3.0\,$\pm$\,1.7   & 2.7\,$\pm$\,0.6         & 2.6\,$\pm$\,0.7 \\
\textsc{bbody} $L$ & 10$^{34}$ erg s$^{-1}$      &  28\,$\pm$\,2             & 13\,$\pm$\,4      & 9.7\,$\pm$\,0.6         & 11\,$\pm$\,3\\

$\chi^2$/dof         &                         &  33/31                  & 137/135  & 201/197 & 145/157\\
\hline
\end{tabular}

\caption{\label{tab:specfits-22june} Spectral  best-fitting parameters
  of  observations taken  on  June  22 (see  the  observations log  in
  Table~\ref{obs-log}).   PC-mode spectra  fitted in  the 0.5--10  keV
  range, WT spectra  in the 2.0--10 keV range. PC  and WT observations
  of ObsID 00700791001 are fitted under the same model.}
\end{table*}

\subsection{XRT timing analysis} \label{sec:timing}
\begin{figure}
\centering
\includegraphics[height=\columnwidth,  angle=-90]{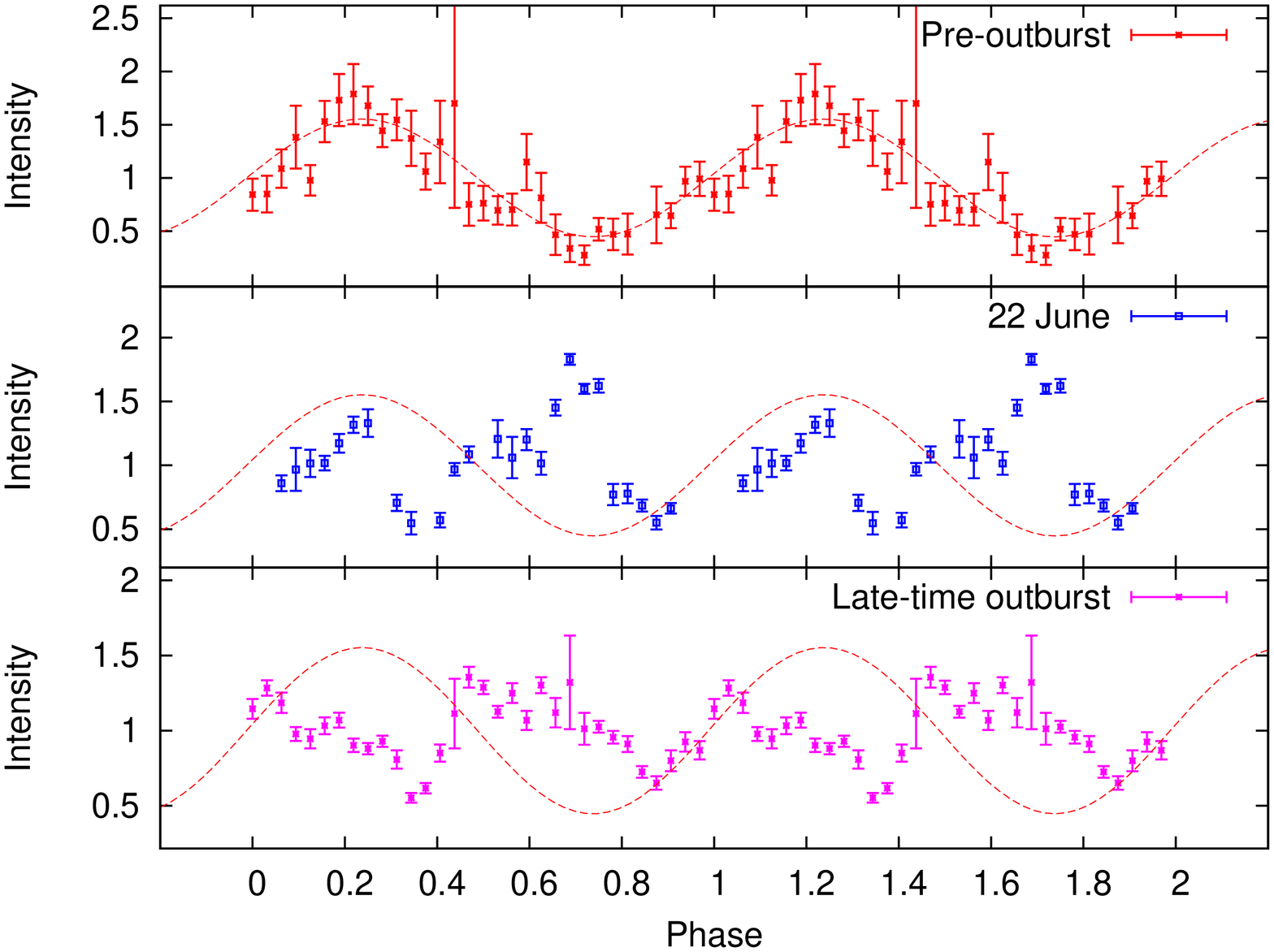}
\caption{Comparison of the 2.0--10.0 keV  folded profiles of \oneE\ at
  the  $P_{\rm  fold}$\,=\,24030.42  s  period  for  the  pre-outburst
  observations (top  panel), the June 22  observations (middle panel),
  and the  late-time post-outburst  observations (bottom  panel).  The
  best-fitting sine  function for  the pre-outburst  data is  shown in
  each panel.}
\label{fig:folding}
\end{figure}

We searched for  periodicities in the \emph{Swift}/XRT  data.  For the
set  of  observations spanning  the  pre-outburst  period, we  clearly
detected  through  folding  search  and also  through  a  Lomb-Scargle
periodogram the  known periodicity  at 6.675 h.  Because of  the short
time frame after the June 22 event, no clear detection was possible in
this time span.  To obtain a time-resolved view of  the changes in the
folded profile,  we choose  $T_{\rm ep}$\,=\,55804.0 MJD  as reference
epoch   and   $P_{\rm   fold}$\,=\,24030.42  s   as   folding   period
\citep{esposito11}.  We  show in Fig.~\ref{fig:folding}  the resulting
folded  profiles using  32 phase  bins. The  pre-outburst profile  can
satisfactorily be  described by a  simple sine function  (upper panel)
with  an amplitude  of  0.50\,$\pm$\,0.03.  The  shape  and the  phase
position of the  maximum agree with the same profile  obtained for the
2006--2011 observations \citep[see fig.~4  in ][]{esposito11}.  In the
middle panel  of Fig.\ref{fig:folding} we  show the pulse  profile for
the June 22 observations (excluding  obsID 00030389032, because of its
relative  small exposure  and exceptional  brightness).  Although  the
folded  profile  is only  sparsely  covered,  the shape  suggests  the
presence of two  antipodal peaks. In the  late-time observations, from
June 23 to July 12, the profile returns to being single peaked, but it
is  significantly more  structured and  with  a clear  shift in  phase
($\sim$\,0.3)  with respect  to  the pre-outburst  profile (see  lower
panel  of Fig.~\ref{fig:folding}).   We also  note a  decrease in  the
pulsed fraction from  $\sim$\,50 to $\sim$\,40 per cent  from the pre-
to the  post-outburst observations.  Finally, we  studied the spectral
hardness as a function of the  phase by choosing as reference hard and
soft bands  the 1--3 and  the 3--10  keV ranges.  We  clearly observed
that the profile peaks are significantly harder, in agreement with the
general pattern already shown in Fig.~\ref{fig:hardness}.

\subsection{UVOT analysis} \label{sec:uvot}
\begin{figure}
\centering
\includegraphics[width=\columnwidth]{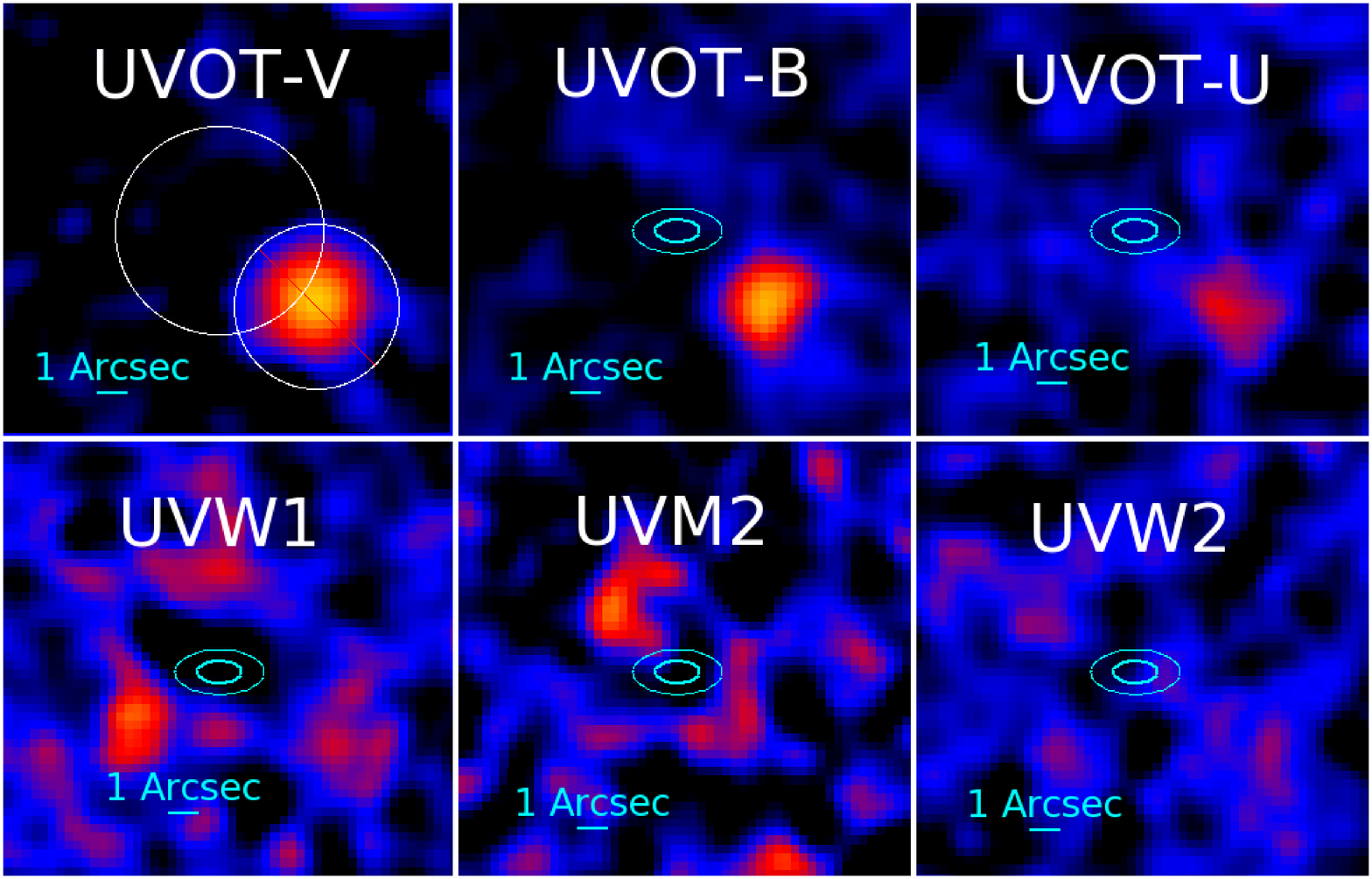}
\caption{Summed  2006--2016 \emph{Swift}/UVOT  exposures in  the three
  optical  and three  UV filters.  The $v$  band panel  shows the  3.5
  arcsec aperture and exclusion region used for the UVOT photometry in
  $v$,$b$, and $u$. In \textit{uvw1} and \textit{uvm2} the source near
  the southern edge was also excluded,  while in $uvw2$ the 3.5 arcsec
  aperture  without exclusion  region  was  used.  The  \emph{Chandra}
  position is  at RA\,=\,244.400958,  Dec.\,=\,--51.040167(J2000), and
  its  68  and   99  per  cent  error  contours  are   based  on  both
  \emph{Chandra} and  VLT data  taken from \citet{deluca08},  and have
  been plotted except in the $v$-band panel to show the absence of any
  detected  emission  at  its  location.  The  crowded  field  appears
  different in different bands.}
\label{fig:uvot}
\end{figure}

The \emph{Swift}/UVOT  began settled observations of  the field around
\oneE~  95 s  after the  BAT  trigger.  During  the first  day of  the
outburst on  June 22, we  collected data  using all UVOT  filters with
various  exposure times  as summarized  in Table~\ref{tab:uvot},  thus
covering the  region in  the 1700--6800 \rm{\AA}  band \citep{kuin15}.
In the summed  images of the $v$, $b$, $u$,  and \textit{uvw1} filters
we detected  a source at a  distance of 3.5 arcsec  which falls partly
within  the aperture  used for  the photometry  (Fig.~\ref{fig:uvot}).
This  source  was   also  reported  in  the  IR  data   south  of  the
\emph{Chandra} position by \citet{deluca08},  therefore, we masked its
emission in all filters (except in the \textit{uvw2} band, where there
is no excess) to derive our upper limits.

Since 2006  $Swift$ has been  monitoring RCW\,103, so we  extended our
analysis to the full available  $Swift$/UVOT data set to derive deeper
limits.  All UVOT data from 2006  to the present time have been summed
and  re-sampled  at   higher  resolution  (see  Table~\ref{tab:uvot}).
Though the UVOT  PSF is only 2.8 arcsec, the  re-sampling of more than
40 exposures brings  out some finer details.  The main  result is that
both  the individual  exposures and  the  summed UVOT  images show  no
evidence for a source at  the \emph{Chandra} position, though possible
sources are  found at distances  larger than 2  arcsec in the  $u$ and
\textit{uvw1} (Fig.~\ref{fig:uvot}).   These were masked  for deriving
the  deeper  UV-optical  limits  at  the  \emph{Chandra}  position  as
reported  in Table~\ref{tab:uvot}.   For a  more immediate  comparison
with the  IR \textit{Hubble Space Telescope}  ($HST$) and \textit{Very
  Large Telescope} ($VLT$) data  analysed in \citet{deluca08}, we also
show   the  error   circles  of   their  \emph{Chandra}   position  in
Fig.~\ref{fig:uvot}.   In the  summed images  of different  filters we
observe  different emission  patterns, including  sources that  have a
significance  just above  3-$\sigma$,  but which  appear  in only  one
filter.  We interpret  this as either being due to  a mixed population
of hot and  cool stars where different stars dominate  the emission in
the  different filters,  or due  to local  variations in  interstellar
extinction.  The  \textit{uvm2} band  is centred on  the 2200~\rm{\AA}
interstellar dust feature,  and some of the differences  seen with the
\textit{uvw1}  and \textit{uvw2}  bands  which have  lower and  higher
wavelengths than the \textit{uvm2} might  be due to differences in the
dust  column.   The   deepest  UV  limits  are   consistent  with  the
conclusions of \citet{deluca08} that a possible binary companion could
only  be of  spectral type  later  than M.   Detected 3$\sigma$  upper
limits  in  the  UVOT  filters  are   shown  in  the  last  column  of
Table~\ref{tab:uvot}.

\subsection{GROND Analysis} \label{sec:grond}

GROND \citep{Greiner2008PASP}  began  observing  the  BAT  error  circle  of
\emph{Swift} trigger  700791 at 02:18:19  \textsc{ut}, after a  short technical
delay, 905 s after the GRB trigger  and 308 s after receipt of the BAT
position. We obtained 45 min of  observations after the trigger, and a
second epoch observation 15 d later.

Reduction and analysis were performed within a custom pipeline calling
upon \textsc{iraf}  tasks \citep{Tody1993ASPC}, following  the methods
described     in     detail      in     \cite{Kruehler2008ApJ}     and
\cite{Yoldas2008AIPC}.  Optical  observations were  calibrated against
tabulated  GROND zeropoints,  whereas the  NIR images  were calibrated
against on-chip  comparison stars from  the Two Micron All  Sky Survey
(2MASS) catalogue \citep{Skrutskie2006AJ}.

\begin{figure*}
\centering
\includegraphics[width=2\columnwidth]{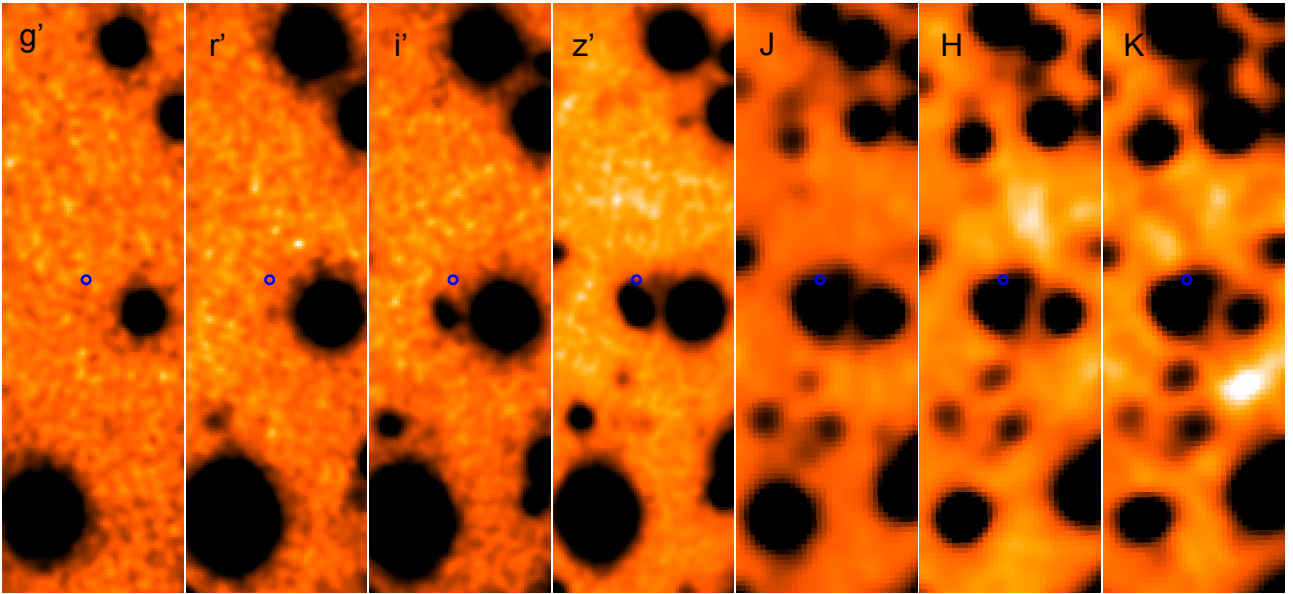}
\caption{Summed   GROND  exposures   in  optical   and  NIR   filters:
  \emph{g$^\prime$},       \emph{r$^\prime$},       \emph{i$^\prime$},
  \emph{z$^\prime$},  \emph{J}, \emph{H},  and \emph{K}  from left  to
  right.  The \emph{Chandra} position is shown by the blue circle.}
\label{fig:grond}
\end{figure*}

Our observations do not reveal  any source inside the \emph{Swift}/XRT
error circle, nor in  the \emph{Chandra} error circle \citep{deluca08}
and  no NIR  flashes at  the \emph{Chandra}  position are  detected in
1-min  images. In  comparison  to  the deep  infrared  images of  that
publication, we detect  the bright complex of multiple  sources to the
south of the error circle starting in the $i^\prime$ band.  We show in
Fig.~\ref{fig:grond} the  field around \oneE, for  each examined band.
Source $2$ in \citet{deluca08} is  detected in our imaging starting in
the $z^\prime$  band.  None of these  sources is seen to  vary between
our observation epochs.  No source is detected at the precise position
of \oneE, especially not the sources  $3$ - $7$ of \cite{deluca08}, in
any  GROND  band at  any  time  with  3$\sigma$  upper limits  on  the
magnitudes (AB system)  for the stacked images of June  22 as reported
in Table~\ref{tab:uvot}.

\begin{table*}
\begin{tabular}{lc rr rr }
\hline
Filter   & Central $\lambda$ & \multicolumn{2}{c}{Exposures (s)} &   \multicolumn{2}{c}{3$\sigma$ upper limits}\\ 
         &\multicolumn{1}{c}{\rm{\AA} (UVOT)/$\micron$ (GROND)}         & 2016-06-22    & 2006--2016   & 2016-06-22  & 2006--2016 \\ 
\hline
\noalign{\smallskip}
\multicolumn{6}{c}{\emph{Swift}/UVOT}\\
\noalign{\smallskip}
\emph{white}    & 3471              & 1227          & 17635       & 22.65 & 24.08\\
\emph{v}        & 5468              & 1629          & 4069        & 20.39 & 20.83\\
\emph{b}        & 4392              & 1082          & 3237        & 21.14 & 21.52\\
\emph{u}        & 3465              & 808           & 30202       & 21.60 & 23.56\\
\emph{uvw1}     & 2600              & 393           & 62597       & 21.38 & 24.51\\
\emph{uvm2}     & 2246              & 2030          & 49138       & 22.66 & 24.69\\
\emph{uvw2}     & 1928              & 6828          & 49177       & 23.46 & 24.53\\
\hline
%\noalign{\smallskip}
%\multicolumn{6}{c}{GROND}\\
\noalign{\smallskip}
\multicolumn{6}{c}{GROND optical bands}\\
\noalign{\smallskip}
\emph{g$^\prime$} & 0.45869&1780.2&$\cdots$ & 23.10& $\cdots$\\
\emph{r$^\prime$} & 0.62198&1780.2&$\cdots$ & 23.10&$\cdots$\\
\emph{i$^\prime$} & 0.76407&1780.2&$\cdots$ & 22.39&$\cdots$\\
\emph{z$^\prime$} & 0.89896&1780.2&$\cdots$ & 22.75&$\cdots$\\
\noalign{\smallskip}
\multicolumn{6}{c}{GROND NIR bands}\\
\noalign{\smallskip}
\emph{J} &1.23992 &960.0&$\cdots$ & 19.79&$\cdots$\\
\emph{H} &1.64684 &840.0&$\cdots$ & 18.91&$\cdots$\\
\emph{K} &2.17055 &960.0&$\cdots$ & 18.70&$\cdots$\\
\hline
\end{tabular}
\caption{\label{tab:uvot}   \emph{Swift}/UVOT   and   GROND   log   of
  observations and  upper limits  for different  NIR/optical/UV bands.
  The magnitudes  (AB system) in the  table are not corrected  for the
  Galactic extinction due to the  unknown, and likely large, reddening
  in the direction of the burst.}
\end{table*}

Using the  best-fitting $N_{\rm  H}$ from the  X-ray spectral  fit and
adopting  the   Galactic  $A_{\rm   V}/N_{\rm  H}$  ratio,   we  infer
$A_{V}$\,=\,5.8  mag.  Using  $A_{K}$\,=\,0.12 $A_{V}$,  our  $K$-band
upper limit corresponds to  flux limit of $<$\,2\,$\times$\,10$^{-13}$
erg   cm$^{-2}$   s$^{-1}$  or   a   $K$-band   luminosity  limit   of
$<$\,3\,$\times$\,10$^{32}$ $D^2_{\rm 3.3 kpc}$ erg s$^{-1}$. Using an
X-ray   flux   at    the   time   of   the    GROND   observation   of
$\sim$3\,$\times$\,10$^{35}$  erg s$^{-1}$  (interpolated between  the
\emph{Swift}/XRT observations), this implies a flux ratio of $f_x/f_K$
$\sim$ 1000.  We note that  the $K$-band flux limits are substantially
smaller than  the $\sim$\,10$^{35}$  erg s$^{-1}$ optical  flares seen
from  Swift  J195509.6+261406,  speculated  to stem  from  a  magnetar
\citep{stefanescu08, castro08}.

\section{Discussion}

We have reported on the characteristics of the \emph{Swift}/BAT bright
X-ray burst coming  from the direction of the RCW  103 nebula observed
on 2016 June 22. At the same time, we have also presented spectral and
timing  results concerning  the X-ray  activation  of the  CCO at  the
centre of the  RCW~103 nebula, \oneE, and the rapid  follow-up in NIR,
optical, and UV bands in our search for a possible counterpart at these
wavelengths.  The long-term  X-ray history of this  source has already
established  flux variations  of about  two orders  of magnitude  on a
years-long time-scale \citep{gotthelf99,  deluca06}. The last outburst
of this source happened between 1999 and 2001, when the source reached
an   intensity   possibly   similar   to   the   one   observed   here
\citep{garmire00}, and  then began a  slow return to  its pre-outburst
luminosity  on   a  years-long   time-scale  \citep[see  fig.~2  in
][]{deluca06}.

This  time we  had the  chance  to closely  monitor the  start of  the
outburst  and  its  short-term  evolution,  because  \emph{Swift}  was
triggered  by the  detection of  an X-ray  burst in  the direction  of
\oneE.  As the characteristics of this burst (duration, spectral shape,
and total fluence)  are typical of soft gamma  repeaters (SGR; whereas
for the same reasons an association with a type-I X-ray burst is ruled
out), and, at the same time,  \oneE\ showed a dramatic change in flux,
spectral shape  and folded  profile, we shall  consider \oneE~  as the
source originating the burst detected by BAT and take this as evidence
for  associating this  peculiar CCO  with the  class of  the magnetars
\citep{duncan92, thompson93, thompson95}.  We note that a very similar
line of  evidence was  sufficient to  grant the  magnetar status  to a
relatively  low  ($B \sim$\,5\,$\times$\,10$^{13}$  G)  magnetic  field
pulsar, PSR J1846--0258, in Kes 75 \citep{gavriil08}.

This  discovery   makes  the  small   group  of  CCO   objects  rather
inhomogeneous based on  the values of their  inferred magnetic fields.
CCO  sources   with  $B$-field   estimates  show  rather   low  values
\citep[$B$\,$\lesssim$\,10$^{11}$     G;     ][]{halpern10,gotthelf13}
compared to typical  values found in young NSs  in high-mass binaries.
\citet{gotthelf08}  coined  the   term  of  \emph{anti-magnetars},  in
antithesis  to the  supercritical  $B$-field values  of magnetars,  to
designate the CCOs hosted in SNRs.  Although \oneE\ was not considered
among the CCO sources listed  in \citet{halpern10} because of its soft
X-ray variability, it  still fulfils all the other criteria  for a CCO
classification.   It   is  most  probably  a   \textit{classical  high
  $B$-field     ($B$\,=\,10$^{14}$--10$^{15}$    G)     )}    magnetar
\citep{deluca06}, and,  even considering a scenario  where the initial
spin-down  was  driven  by  an ejector  phase  of  \textit{magnetized}
debris, the  required dipole  field would still  be above  10$^{12}$ G
\citep{ikhsanov13}.   This  suggests  that  it  can  be  difficult  to
generalize and assume all CCO objects as young and very low magnetised
NSs    \citep[see   e.g.,][for    magnetars    hosted    in   a    SNR
  environment]{gaensler01, vink08,gao16}, unless  very ad hoc criteria
are chosen.

We studied the soft X-ray evolution of the source in the first 3 weeks
of the  outburst thanks  to the  monitoring campaign  of \emph{Swift}.
The X-ray light curve shows a clear peak just close to the time of the
BAT trigger,  with a steep  decrease in  the following hours,  until a
plateau is reached within 1 d from the X-ray burst. The flux evolution
in  the following  weeks  did not  show any  evident  sign of  fading,
suggesting, as in the previous outburst,  a possible slow decay to the
pre-outburst  luminosity  levels. The  initial  steep  decay, and  the
flatter evolution  is similar to  what observed  in the case  of other
magnetars outbursts \citep[see e.g. the decay of the SGR 1E 2259+58 in
  its 2002  June outburst; ][]{woods04}, and,  more generally, closely
resembles the behaviour of transient  magnetars like the SGR 1627--41,
that   shows   similar   flux  variations   on   similar   time-scales
\citep{esposito08}.  We tracked the  most significant spectral changes
using as a benchmark  the time-averaged \textit{quiescent} spectrum of
\oneE\ from the  \emph{Swift}/XRT observations taken about  2 yr prior
to 2016 June 22. We did  not choose to disentangle the \oneE\ emission
from the contribution of its nebula, because of the intrinsic bias and
dependence of the results from the choice of the source and background
extraction regions.  Instead we modelled  both components in  a single
fit to the data, using the results  from the extensive work on the SNR
emission  made  by  \citet{frank15}.   In  this  way,  we  obtained  a
statistically  acceptable  description  of  the  data,  and  we  could
constrain  at much  higher confidence  the parameters  determining the
spectral  state of  the source.   The  \oneE\ emission  along all  the
outburst showed little variation  in the time-averaged spectral shape,
characterized by  a soft thermal component  of temperature $\sim$\,0.6
keV and  a hard X-ray  tail, carrying about  10\% of the  total source
emission.  The spectral shape in the  first observation of June 22, 20
min earlier  than the  BAT burst,  seems to be  harder than  the other
late-time spectra, but we also  note that observations performed a few
hours after the BAT event showed  a rapid return to the temperature of
$\sim$\,0.6  keV that  also characterized  the pre-outburst  spectrum.
The   outlined  spectral   characteristics   such   as  peak   thermal
temperature,  harder flux  excess during  the outburst,  time-scale of
flux variations are all in agreement with the general properties shown
by        transient        magnetars       \citetext{see        e.g.\,
  \citealp{kaspi03,scholz11,scholz12},   or  these   general  reviews:
  \citealp{rea11, mereghetti15, turolla15}}.

We studied the timing characteristics  of the pulsed profile of \oneE,
comparing the profiles  at different times. We remark  that, given the
short time-span of  the observations during the outburst,  we were not
able to clearly detect the 6.67  h periodicity, however, it is evident
that any reasonable change in  its value cannot have any statistically
significant effect on the folded profile.

A sort  of bimodality in  the pulsed profile  was shown by  the sparse
observations  of this  source in  the 1999--2005  years, where  it was
already found that when \oneE\ was in a brighter state the profile was
remarkably different,  and more structured \citep{deluca06}.   We have
observed that  this change is  not gradual, but  it happens on  a very
short    time-scale   at    the    time   of    the   outburst    peak
(Fig.~\ref{fig:folding}). The folded profile in outburst clearly shows
that a  significant phase  change took place,  and similarly  to other
magnetars     where    the     same    phenomenology     is    present
\citep{kaspi03,woods04,dib09,woods11},  it  could indicate  a  general
re-arrangement  of the  magnetic field,  which also  caused the  rapid
dissipation of energy in the burst event.

Clearly,  it is  comparatively much  more difficult  to assess  if the
sudden  change  in  the  folded  profile is  also  associated  with  a
frequency glitch, as observed in  many magnetars \citep{dib08}, as the
fractional  frequency shifts  are  generally  less than  a  part in  a
million  except in  some exceptional  cases \citep{palmer02}.   Future
observations,  spanning  a longer  time-frame,  will  hopefully set  a
constraint on  this issue.  Because  of this characteristic  change in
the  pulsed profile,  commonly  observed after  a  burst in  magnetars
\citep{mereghetti15}, we  believe that the 6.675  h periodicity cannot
be of orbital origin as  had being speculated in \citet{bhadkamkar09},
but it must be associated with  the NS spin period.  Early suggestions
for the presence of $dips$  in the folded profile \citep{becker02} are
ruled out, as the spectral  hardening appears strongly correlated with
the total flux over the entire flux  range, and it is not localized in
the  bottoms of  the folded  profile (Fig.~\ref{fig:hardness}).   This
finding makes  \oneE\ the slowest  pulsar to our  knowledge \citep[the
  second  being  RX  J0146.9+6121,  with  a spin  period  of  1380  s;
][]{haberl98}, and also makes \oneE\ a rather exceptional object among
all the known  magnetars, because the distribution of  spin periods of
these objects lies  in only a decade of periods  between $\sim$\,2 and
12 s \citep{olausen14}. Because of this extreme slow spin, the present
rotational    energy   stored    in   the    NS   would    be   $\sim$
3.4\,$\times$\,10$^{37}$ erg  (assuming a canonical moment  of inertia
10$^{45}$ g  cm$^{2}$), and  this value  is very  close to  the energy
dissipated  in the  burst event  ($\sim$\,2\,$\times$\,10$^{37}$ erg),
thus indicating that magnetic dissipation  of an intense field must be
the main,  if not the only  source responsible for the  observed burst
radiation.

If the spin period of the NS at its birth was similar to that of other
magnetars,  some mechanism  must have  furiously braked  it down  on a
time-scale comparable with the SNR age,  which is only $\sim$ 2000 yr.
Within  the  magnetar  scenario, \citet{deluca06}  proposed  that  the
braking could be provided by a  propeller effect due to a reservoir of
mass formed from  the SNR material fallback \citep[a  fossil disc, see
  also][]{wang06}, continuously expelled  at the magnetospheric radius
of the NS.   The only constraint which appears reasonable  is that the
NS initial period should have been longer than 0.3 s to avoid the disc
disruption by the relativistic outflow  of the newly born active radio
pulsar.       A      magnetar      with     a      magnetic      field
$B$\,=\,5\,$\times$\,10$^{15}$ G  could reach  the period  observed in
\oneE,  after having  expelled 3\,$\times$\,10$^{-5}$  M$_{\odot}$, in
less  than  the age  of  the  SNR.  It  is  interesting  to note  that
\citet{reynoso04}  found the  presence  of  an H\textsc{i}  depression
region around \oneE\ of radius 64 arcsec, and a lack of evidence for a
possible ionized H\textsc{ii} region.   The missing mass was evaluated
to be $\sim$\,0.3 M$_{\odot}$, thus  suggesting that a strong sweeping
of  material  at  the  centre  of the  SNR  might  have  taken  place.
\citet{li07} has further explored this  scenario through a Monte Carlo
simulation of a population of 10$^{6}$ NS magnetars interacting with a
fallback disc.   The NS  population differs  in initial  spin periods,
axis orientations, $B$-field, and mass of the fallback disc.  He found
that most of  the magnetars ($\sim$\,99 per cent) would  be found 2500
yr after their birth in the ejector phase (when the radiative pressure
from the NS keeps the surrounding  plasma away from the light cylinder
and the spin-down  can only be provided by  magnetic dipole emission),
but that  0.6\% could be found  in the propeller phase  (when the disc
radius is  between the magnetospheric  and the light  cylinder radius)
and  be effectively  braked  to periods  $>$\,10$^{3}$  s as  possibly
happened for \oneE.

Alternatively, the  \oneE\ could be a  binary system formed by  a very
low mass star and a magnetar  with a spin (quasi-)synchronous with the
orbital  period  \citep{pizzolato08}.   In this  scenario  the  torque
needed to slow down the NS is provided by the interactions between the
magnetic field  and the surrounding  material, similar to the  case of
white  dwarfs  in intermediate  polars.  However,  the presence  of  a
low-mass companion which could  have survived the supernova explosion,
is rather  unlikely, given that  such mass-lopsided system  would have
been  prone  to unbinding.   In  this  context,  it is  relevant  that
\citet{deluca08}  ruled out  all  but late  M-type  stars as  possible
companions,  while  \citet{li07} showed  that,  even  in the  case  of
survival, an irradiation-induced  wind would not be able  to power the
observed X-ray emission.  In search  of possible counterparts at other
wavelengths, we have  also shown the results from  a rapid optical/NIR
follow-up of \oneE\  with GROND made 0.26 h after  the BAT trigger and
by \emph{Swift}/UVOT in  the optical/UV bands both at the  time of the
trigger  and in  the whole  set  of \emph{Swift}  observations of  the
source.  No significant counterpart was detected in the stacked images
consistent with the  position of \oneE.  We derived a  series of upper
limits on the  magnitudes in the different bands from  NIR to UV, thus
supporting  the absence  of any  irradiated close  companion star,  or
distant accretion flow \citep{wang07}.

\section{Acknowledgements} 
The authors are  very grateful to Amy  Y. Lien for her  support in the
data reduction and analysis of the BAT data.\\ PAE, APB, NPMK, and JPO
acknowledge  \emph{Swift} funding  from  the UK  Space  Agency. \\  We
acknowledge contract  ASI-INAF I/004/11/0.\\  Part of the  funding for
GROND (both hardware as well as personnel) was generously granted from
the Leibniz-Prize to  Professor G. Hasinger (DFG  grant HA 1850/28-1).
This  work made  use of  data supplied  by the  UK Swift  Science Data
Centre at the University of Leicester.\\ This research has made use of
the  XRT   Data  Analysis   Software  (XRTDAS)  developed   under  the
responsibility  of   the  ASI  Science  Data   Center  (ASDC),  Italy.
\bibliographystyle{mn2e} 
\bibliography{refs}

%%%%%%%%%%%%%%%%%%%%%%%
%%%%%%%%%%%%%%%%%%%%%%%%%%%%%%%%%%%%%%%%%%%%%%%%%%
\end{document}